\documentclass[acus]{jac2003}
\usepackage[colorlinks]{hyperref}%
\usepackage{cite}
\usepackage{graphicx}
\usepackage{epsfig}

\catcode`@=11 
\newdimen\z@ \z@=0pt 
\newskip\z@skip \z@skip=0pt plus0pt minus0pt
\def\m@th{\mathsurround=\z@}
\def\ialign{\everycr{}\tabskip\z@skip\halign} 
\def\eqalign#1{\null\,\vcenter{\openup\jot\m@th
  \ialign{\strut\hfil$\displaystyle{##}$&$\displaystyle{{}##}$\hfil
      \crcr#1\crcr}}\,}
\catcode`@=12 
\let\cl=\centerline
\def\figbox#1;#2;{\parbox{#2cm}{\epsfig{file=\figdir #1.eps,width=#2cm}}}
\def\figboxc#1;#2;{\cl{\figbox#1;#2;}\vglue-2mm}

\def\ie{{\it i.\kern-.5pt e.\kern1pt}}  \def\etal{{\it et al.}}
  
\def\up#1{$^{#1}$}  \def\dn#1{$_{#1}$}
\def\ifm#1{\relax\ifmmode#1\else$#1$\fi}
\def\Bbar{\ifm{\rlap{\kern.22em\raise1.9ex\hbox to.58em{\hrulefill}} B}}
   \def\deg{\ifm{^\circ}}

\def\to{\ifm{\rightarrow}} \def\sig{\ifm{\sigma}}   
\def\K{\ifm{K}}  
\def\ff{$\phi$--factory}  \def\DAF{DA$\Phi$NE}  \def\f{\ifm{\phi}} 
 \def\pic{\ifm{\pi^+\pi^-}} \def\pio{\ifm{\pi^0\pi^0}} 
  
\def\po{\ifm{\pi^0}}
\def\ks{\ifm{K_S}} \def\kl{\ifm{K_L}} 
  \def\ksl{\ifm{K_{S,\,L}}}
\def\eps{\ifm{\epsilon}} \def\epm{\ifm{e^+e^-}}
\def\rep{\ifm{\Re(\eps'/\eps)}}    
\def\Kb{\ifm{\rlap{\kern.3em\raise1.9ex\hbox to.6em{\hrulefill}} K}}
  
\def\C{\ifm{C}}  \def\P{\ifm{P}}  \def\T{\ifm{T}}
\def\noc{\relax\hglue0pt{\rlap{$C$}\raise.15ex\hbox{$\kern
.18em\backslash$}}}
\def\nop{\relax\hglue0pt{\rlap{$P$}\raise.15ex\hbox{$\kern
.18em\backslash$}}}
\def\noT{\relax\hglue0pt{\rlap{$T$}\raise.15ex\hbox{$\kern
.18em\backslash$}}}
\def\nocp{\noc\nop} \def\nocpt{\noc\nop\noT}
\def\ko{\ifm{K^0}}  \def\kob{\ifm{\Kb\vphantom{K}^0}}
\def\gam{\ifm{\gamma}}  
 \def\ab{\ifm{\sim}}  \def\x{\ifm{\times}}
\def\sta#1{\ifm{|\,#1\,\rangle}} 
\def\L{\ifm{{\cal L}}}  
\def\pt#1,#2,{\ifm{#1\x10^{#2}}}
  
  \def\dif{\hbox{d}}   
\def\epss{\ifm{\eps_S}}  \def\epsl{\ifm{\eps_L}}
\def\bye{\end{document}}
\def\figdir{}

\begin{document}
\pagestyle{myheadings}
\markboth{}{{\rm Closing Remarks at the {\it Workshop on $e^+e^-$ in the 1-2 GeV Range}, Alghero, 10-13 September 2003}}
\thispagestyle{empty}
\title{\vglue-.5cm Physics at \DAF2\\
{\rm Closing Remarks at the Alghero \DAF2 Workshop, 10-13 September, 2003}}

\author{Paolo Franzini\\Universit\`a di Roma, {\it La Sapienza}}

\maketitle

\begin{abstract}
\noindent The original plan of a vigorous program of frontier physics at LNF, based on the \ff\ \DAF\ with the KLOE detector, has not yet reached its major aim in the study of discrete symmetries in the neutral kaon system. It has however led to the education of a new generation of particle and accelerator physicists. With such invaluable human resources it seems appropriate to consider a renewed effort in achieving much improved collider performance. It will be  argued in the following that it is still possible to envision a program of superior quality physics, requiring several years and broad enough to justify a new collider operating at the \f-meson mass. It remains of paramount importance that the program covers topics of fundamental interest, with many collateral avenues well connected to the ultimate goal, a most sensitive test of \C\P\T\ invariance. The appropriate time frame for these ventures to be successful is quite well defined, in view of the large LHC effort to begin towards the end of the present decade.
\end{abstract}

\section{\DAF2, what is it?}\noindent
``Collider-factories'' have done quite well in some laboratories, SLAC and KEK, but not quite so at LNF~\cite{slakek}. On the basis of recent ideas, ``super collider-factories'' are being considered~\cite{cb}. In the following I will discus the possibilities of doing superior physics at a collider operating at the \f-meson mass, \ie\ at a center of mass energy $W$=1019 MeV, with a luminosity \L=\pt5,34, cm\up{-2} s\up{-1} or, in less wordy units, 50 nb\up{-1}/s. It is most convenient to think in terms of $\mu$b\up{-1}/s, which corresponds to the production of three \f-mesons per second, three charged kaons per second and one \ks- and \kl-mesons per second.

Whatever the program might be, it must be considered in the context of other current projects in the particle physics community. We are at a sort of in between times, a period during which a few activities are continuing, after the excitements of the past decade, in preparation for the LHC era. The Tevatron program is potentially interesting but is plagued by many problems, HERA is winding down, KEK and SLAC are heavily involved in unravelling an unwieldy Gordian knot (bundle?).

That is just the right time for us to bring \DAF\ back to its original goal of measuring all the parameters in the kaon sector, but also well beyond it, in its ultimate accuracy, as is appropriate for 10 years later. Remember that while the double ratio
$${\Gamma(\kl\to\pic)\over\Gamma(\ks\to\pic)}\left/{\Gamma(\kl\to\pio)\over \Gamma(\ks\to\pio)}\right.$$
is well measured, partial widths and amplitude ratios are much more poorly known.
\subsection{The Laboratori Nazionali di Frascati, LNF}\noindent
LNF has made a major investment in \DAF~\cite{dafne}; and most of the infrastructure remain fully adequate to a new phase of endeavors. The upgraded collider's footprint remains essentially unchanged; so does that of the detector, refs. \citen{klopro}-\citen{klotri}.
Infrastructures are there; from real estate and management to utilities (water? no sewer problem).
LNF is the only INFN laboratory that has an accelerator division that built a working and productive collider. Furthermore, it's the birthplace of colliders after all~\cite{ada}.
Without anyone much noticing, a miracle has occurred in the last dozen of years around LNF. A new generation of physicists has emerged from the junior staff of LNF and the collaborating institutions (like the travelling minstrels).
They can THINK, DO, ANALYZE, be RESPONSIBLE, on their own AND, even more importantly, work devotedly as a COHERENT team towards a COMMON end.\\ 
They are, without any doubt, world class in their skills and motivation, and I'm extremely proud of them. Let's give them an equally worthy instrument to work with!

It is absolutely necessary to the vitality of the laboratory to have a new project.
It is an IDEAL project for INFN and helps to maintain its budget justification. It also satisfies many of the following relevant points. It has a challenging, though yet to be proven feasible, design. It is a complete and self contained project, not a piece of some international humongous project in which INFN is a small part.
When built, it will have international visibility, as the present KLOE result are beginning to have, refs. \citen{klossl}-\citen{kloetaggg}. \DAF 2 certainly would have the UNIQUENESS of being the only BRIGHT phi factory (VEPP-2000's goal luminosity is orders of magnitude lower).
The project fits well in the temporal period indicated above, including completion of KLOE's present physics program and preparing a realistic machine and detector design. The upgraded KLOE would still have the largest chamber, the fastest calorimeter, plus more tracking close to the interaction point and a good Q-cal.
\section{PHYSICS}
\begin{figure}[htb]
\centering
\includegraphics*[width=58mm]{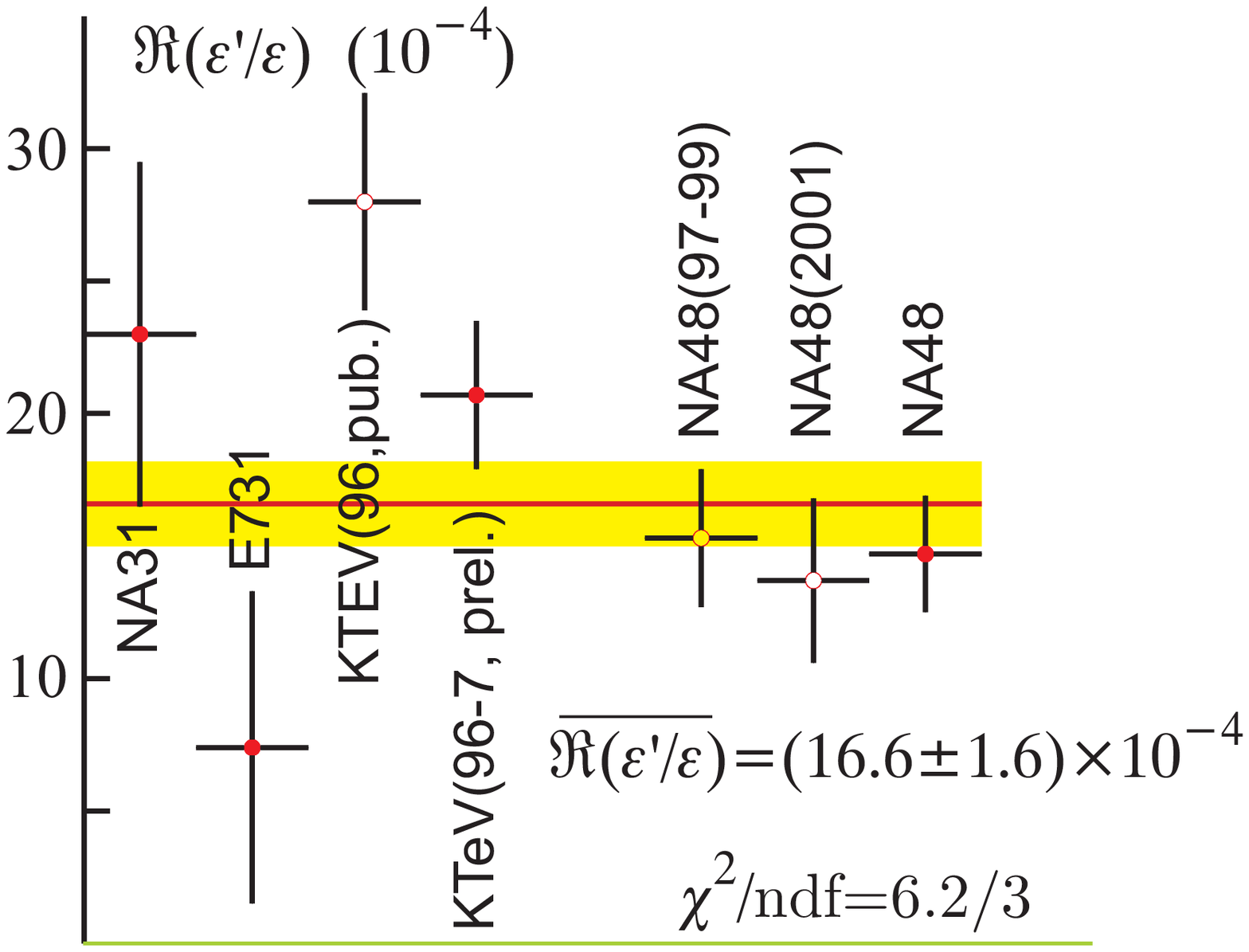}
\caption{Results for \rep\ from CERN and FNAL, from Lenti, CERN seminar.}
\includegraphics*[width=70mm]{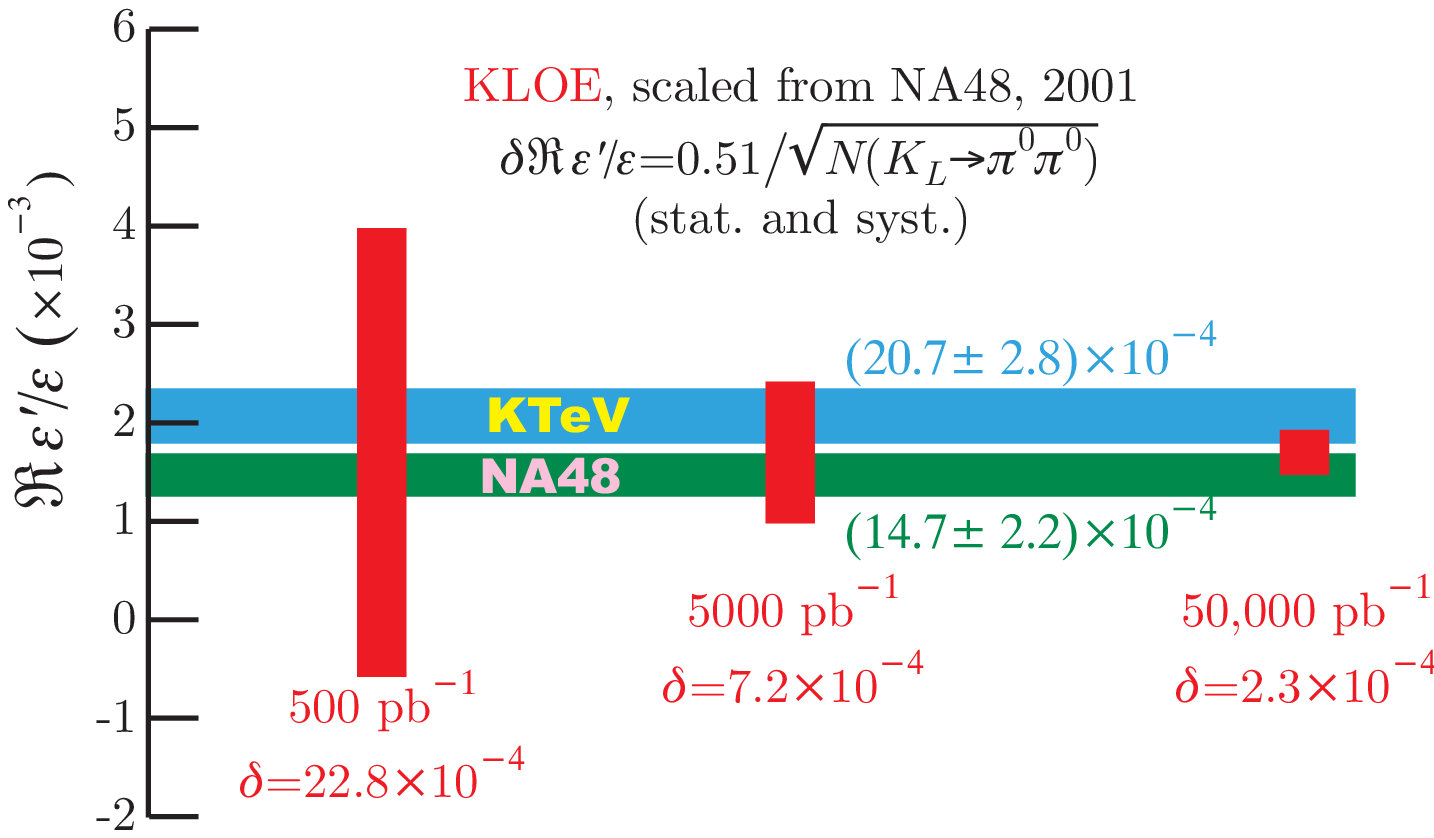}
\caption{Comparison between the results above and the possible KLOE contribution far various value of the accumulated luminosity..}
\label{fig:rep}
\end{figure}

\noindent
The original KLOE proposal~\cite{klopro} was centered around proving the existence, or otherwise, of direct \nocp. While we proposed to do this by measuring the four rates
$$\Gamma(\ksl\to\pic,\pio)$$
we did emphasize the uniqueness of a \ff\ in providing interferometry~\cite{pf}, thus allowing the measurement of phases and magnitudes of the amplitude ratios
$$\eta_i={A(\kl\to i)\over A(\ks\to i)}$$
as well as {\it kinematical} properties such as $\Gamma_{S\!,\,L}$ and $\Delta m$. Only in this way it is possible to measure almost all the parameters of the neutral kaon system, see ref.~\cite{maia}.
Direct \nocp\ has meanwhile been proven by NA48~\cite{nafour} and KTeV~\cite{ktev}, while KLOE has not much to say yet, fig. \ref{fig:rep}. Contrary to the expectations of a few years ago, there is a consensus today that there is no way to reliably connect \rep\ and the CKM parameters. Moreover this might not change for some time to come. One could then argue that there is little reason for spending time and money in trying to perform a third measurement.
However, measuring the $\eta_i$ parameters (and $\Gamma$'s and more) remains a fundamental job to be performed, in order to complete our knowledge of the parameters of the neutral kaon system. And this alone is already quite a justification for \DAF2-KLOE.

\begin{figure}[htb]
\centering
\includegraphics*[width=70mm]{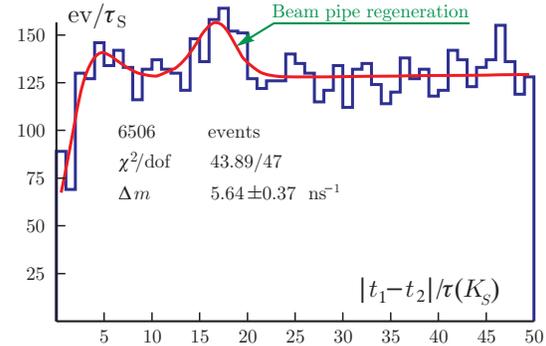}
\caption{The first example of interference, for \f\to\ks\kl\to\pic,\pic, ever  observed. KLOE, 2003. The interference pattern would allow measuring $\Delta m$ to a better accuracy than presently known, with the \DAF2 luminosity.}
\label{fig:interf}
\end{figure}

\begin{figure}[htb]
\centering
\includegraphics*[width=60mm]{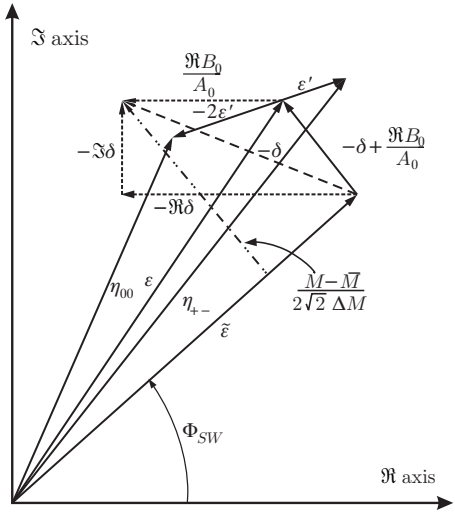}
\caption{Relation between the $\eta$ ratios, $\eps,\; \eps'$ as well as \C\P\T-violating quantities.}
\label{fig:etarho}
\end{figure}

This becomes truly interesting at a super-factory, where fine measurements of all the parameters describing the neutral kaon system become possible, because of the unique possibility, and this is true only at a \ff, of measuring amplitudes and phases of all relevant amplitudes. One can note here that the coherence property of the two kaon wave function has been known and expected for more than 50 years. 
KLOE for the first time, with very little data unfortunately, has been able in fact to observe interference, as shown in fig. \ref{fig:interf}. I remind you that these measurements allow measuring phases through the appearance of the term below in the time difference distribution.
$$\eqalign{&I(f_1,f_2,\Delta t)=\cr
&\qquad\dots2|\eta_1||\eta_2|e^{-\Gamma\Delta t/2}\cos(\Delta m\Delta t+\f_1-\f_2)\cr}$$
Measuring the complex amplitude ratios $\eta_i$ means in fact measuring \rep\ and much more. From the relations
$$\eqalign{\eta_{\pic}&=\eps+\eps'\cr
           \eta_{\pio}&=\eps-2\eps'\cr}$$
(which could be taken as the definition of \eps\ and $\eps'$) one obtains
$$\eqalign{\eps&=(2\eta_{+-}+\eta_{00})/3\cr
           \eps'&=(2\eta_{+-}-\eta_{00})/3\cr}$$
which provide much more information than just \rep. The interconnection is indicated in fig. \ref{fig:etarho}, which is drawn including possible \C\P\T-violation effects: $\delta,\;M_K-M_{\overline K}\ne0$~\cite{pfa}.

\section{An aside: KLOE, 99-03}
\noindent
In spite of a large amount of frustration, KLOE has made fundamental contributions to the study of:\\
1. \ks\ decays, rare and not so rare\\
2. Scalar and pseudoscalar mesons\\
3. \sig(\epm\to\pic), of relevance to the muon anomaly.

A \ff\ is unique for \ks\ study. Only from \f-decays we can get pure \ks\ (and \kl\ and \K\up+ and \K\up-) beams. Yields are ${\cal O}$(10\up6)/pb\up{-1} kaons of any kind. After tag and fiducial volume one is left with 10-50\% of them. Purity is unsurpassed and (not often appreciated) an {\it absolute count} is automatic.
In the 2002-3 edition of PDG, KLOE appears for the first time, with 11 entries. And all the measurements are already vastly improved in our newer results. By the end of 2003, we will provide the basis for the first improvement, in a long time, $>30$ years, of the $|V_{us}|$ value. And, for the first time, our data will allow critical checks of chiral perturbation calculations.

Still the best product of KLOE are all the young people who have had the opportunity to struggle and solve lots of problems to get to the final results.
\subsection{Aside on errors}
\begin{figure}[htb]
\centering
\includegraphics*[width=70mm]{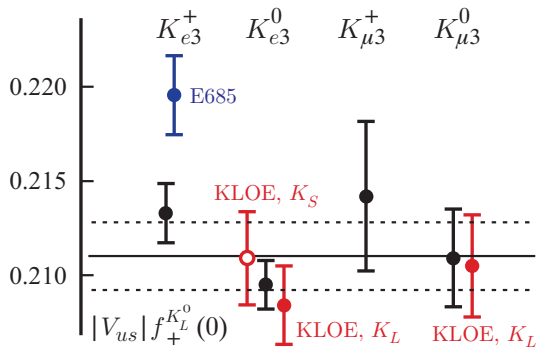}
\caption{$|V_{us}|\x f_+^0$ from kaon semileptonic decays for 5 different modes, using the PDG error.}
\label{fig:vus}
\end{figure}

\begin{figure}[htb]
\centering
\includegraphics*[width=70mm]{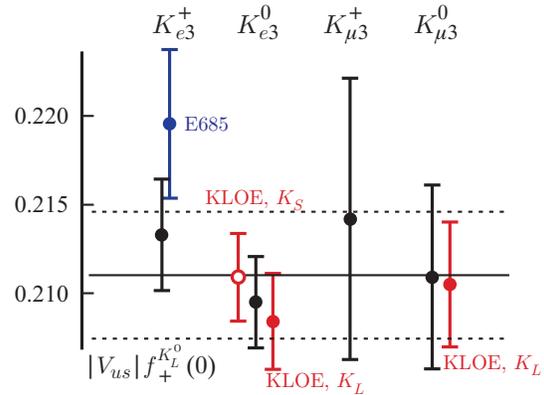}
\caption{Same as fig. \ref{fig:vus}, but with  possibly more realistic errors.}
\label{fig:vusa}
\end{figure}

\noindent
If you search in the Review of Particle Physics~\cite{pdg} by the PDG for the data used to get $|V_{us}|$, you
{\it first} realize that they come mostly from 1972 and earlier. One exception is $\tau(K^\pm)$ which was last measured in 95, with poor agreement between the two result of the same experiment. Given the existing data, I would conclude that the lifetime error is 0.8\%, rather then the quoted 0.2\%. {\it Secondly}, you notice that the branching ratio errors come from the PDG fit, the actual measurements have much large errors. So the recent plot of $|V_{us}|\x f_+^{\ko}$, which is shown in fig. \ref{fig:vus}~\cite{vuszz}, might in fact be more like fig. \ref{fig:vusa}.
Note that the decay \ks\to$\pi^\pm$$e^\mp$$\nu(\bar\nu)$, unobserved until 1999, appears in the plot with approximately the same accuracy as \kl\ and charged kaons.
A problem with the PDG fits (with all fits) is that they give smaller errors and many, large, correlations.

\subsection{Uniqueness of a \ff}\noindent
A \ff\ is unsurpassed in providing \ks-\kl\ and $K^+$-$K^-$ pairs in precisely prepared states, with low background as summarized in the table.\vglue3mm

\def\dv{\vglue.8mm}\begin{tabular}{|c|c|}
\hline
Background & \parbox{4cm}{\centering\dv\sig(\epm\to\f)$\gg$cont., $>\,$Bhabha (large $\theta$)\dv} \\ \hline
Purity   & \parbox{4.5cm}{\centering\dv$\psi(0)=(\ks\kl-\kl\ks)/\sqrt2$\\ $K^\pm$!!\dv}  \\ \hline
         &\parbox{4.5cm}{\centering\dv$K^+K^-$: 50\%\dv}  \\
Yields   & \parbox{4.5cm}{\centering\dv\ks\kl: 34\%\dv}\\
         & \parbox{4.5cm}{\centering\pic\po: 15\%\dv}  \\ \hline
\parbox{1.3cm}{\centering\dv$\delta p/p$\dv} &  0.5\%, from machine $\delta E$ \\ \hline
 $\beta(\ko)$ &  \parbox{4cm}{\centering\dv\ab0.2; $\delta\beta/\beta$\ab0.5\%, from machine $\delta E$\dv}\\ \hline 
\end{tabular}
\subsection{KLOE Results: Masses}
\begin{figure}[htb]
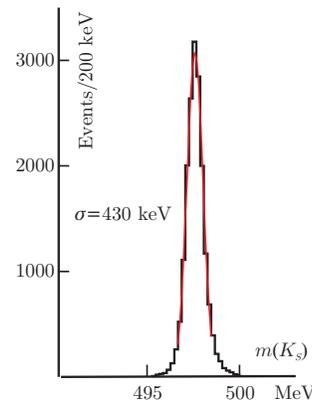

\centering
\figboxc mksant;4;
\caption{KLOE \ks\ mass resolution.}\vglue3mm
\label{fig:kloemass}
\end{figure}

\begin{figure}[htb]
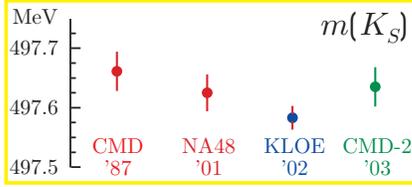

\centering
\figboxc mks-all;5.5;
\caption{Recent results for $m(K)$.}\vglue3mm
\label{fig:kmass}
\end{figure}

The ideal kinematics allows precise mass measurements. Figure \ref{fig:kloemass} shows the \ko\ KLOE mass resolution
Recent high precision \ko\ mass results are shown in fig. \ref{fig:kmass}. We find $m(\ks)=497.583\pm0.005\pm0.020$ MeV. The result proves the accuracy of the KLOE momentum scale. Radiative corrections, both at \DAF\ and VEPP-2M, are the main uncertainties, refs. \citen{cmdkm}-\citen{cmd-2km}.
\subsection{KLOE Results: \ks\ semileptonic decays.}
\noindent
KLOE has collected ${\cal O}$(10\up4) semileptonic \ks\ decays~\cite{klossl}. A partial sample signal is shown in figures \ref{fig:ksslp} and \ref{fig:ksslm}.
\begin{figure}[htb]
\centering
\figboxc ksslplus;3.7;
\caption{\ks\ semileptonic decay signal: positrons.}\vglue3mm
\label{fig:ksslp}
\figboxc ksslminus;3.7;
\caption{\ks\ semileptonic decay signal: electrons.}\vglue3mm
\label{fig:ksslm}
\end{figure}
Our preliminary resulta are: 1. BR(\ks\to$\pi e\nu$)=\pt(6.9\pm0.15),-4, or  $\delta\Gamma/\Gamma$=2.2\%; 2. ${\cal A}^e_S$=\pt(19\pm18),-3,; 3. $\Re x=0.003\pm0.0065$, or $\Re x<1.3\%$. With all '02 data we will reach an accuracy of 0.5\%. These results demonstrate the power of KLOE's \ks\ tagging and particle ID.
\subsection{KLOE Results: \ks\to\pic(\gam)/\ks\to\pio}\noindent
The ratio $R=\Gamma(\ks\to\pic)/\Gamma(\ks\to\pio)$ has been measured for over 50 years. Its value of \ab2 lead to postulating the $\Delta I$=1/2 rule, which appears to be universally valid in all $|\Delta S|$=1/2 decays of strange particles.
\begin{figure}[htb]
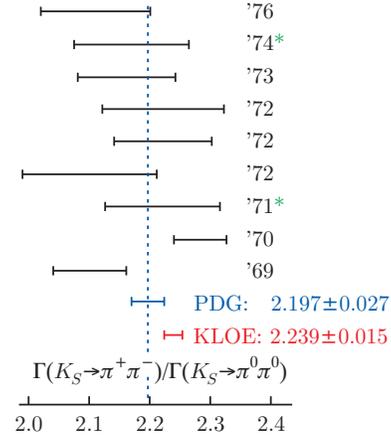

\centering
\figboxc ksppbr;5;
\caption{$R$ in the last 33 years.}
\label{fig:picpio}
\end{figure}
The suppression of $|\Delta S|$=3/2 (or enhancement of $|\Delta S|$=1/2) transition is still unexplained. KLOE for the first time has reached a statistical accuracy of 0.15\% and has properly accounted for radiative effects. This result was obtained with a trigger efficiency $\geq>$96.5\% and an overall acceptance of \ab57\%. We find $R=2.239\pm0.003 ({\rm stat.})\pm0.015{(\rm syst.)}.$~\cite{klopipibr}
An error $\delta R/R=0.1\%$ contributes \pt1.6,-4, to the error on \rep.
KLOE has collected enough data to reach an accuracy of \ab0.1\% in $R$. The preliminary results obtained with \ab5\% of the total statistics are again proof of the value of tagging and of the use of auxiliary data samples to determine efficiencies and acceptances. All the value mentioned above where obtained in this way. Monte Carlo simulation are used to perform integrations over geometrical boundaries and to confirm the results of the analysis. Direct results are quite more believable than grand fits to heterogenous information. 
\subsection{KLOE Results: Scalars and Pseudoscalar\\ Mesons}\noindent
For a long time the lightest scalar $q\bar q$ states have been a hard to understand puzzle. Radiative transitions are a very effective tool for the study of quark structures. At \DAF\ scalar mesons are ideally studied via the reactions \f\to$f_0$\gam\ and \f\to$a_0$\gam.
\begin{figure}[htb]
\centering
\figboxc popogam;5;
\caption{$a_0$ signal in KLOE.}\vglue3mm
\figboxc etapogam1;5;
\caption{$f_0$ signal in KLOE.}\vglue3mm
\figboxc picgam1;4.8;
\caption{First evidence for \f\to$f_0$\gam\to\pic\gam.}\vglue3mm
\label{fig:fopic}
\figboxc etap2;5;
\caption{$\eta'$ signal in KLOE, from all data. There are \ab500 events in the $\eta' peak.$}
\label{fig:etap}
\end{figure}
From early analyses we found BR(\f\to\pio\gam)=\pt(1.09\pm0.06),-4,, BR(\f\to$\eta$\po\gam)=\pt(0.85\pm0.08),-4,, BR($\f\to\eta'\gam$)=\pt6.10\pm0.7,-5,, refs. \citen{klofnot} and \citen{kloa-o}. These results provide the first clean look at the structure of the scalar mesons. 
The first accurate measurement of singlet-octet mixing as well as limits to the gluon contents of $\eta-\eta'$, comes from the study of $\f\to\eta'\gam$. From the small 2000 sample we find  BR($\f\to\eta'\gam$)=\pt6.10\pm0.7,-5,, ref. \citen{kloetap}. Fig. \ref{fig:etap} shows the present KLOE $\eta'$ signal.
\subsection{KLOE Results: \sig(\epm\to\pic)}\noindent
The recent accurate measurements of the muon anomaly in Brookhaven have brought attention to the necessity of improved measurements  of the cross section for \epm\to hadrons, an important part of which is the process \epm\to\pic\ from to threshold to about 1 GeV.
\begin{figure}[htb]
\centering
\figboxc sigppg03;5;\vglue5mm
\figboxc pipiff;5;
\caption{$\dif\sig(\epm\to\pic\gam)/\dif M^2_{\pic}$ (top) and $|F_\pi(M^2_{\pic})|^2$ \  (bottom).}
\label{fig:sighad}
\end{figure}
Traditionally this measurements have been done by changing the collider energy, a difficult process, requiring also many auxiliary measurement. With a high quality detector and adequate luminosity, it becomes convenient to let initial state radiation, ISR, provide variable energy for hadro-production up to the collider energy. The KLOE results, see ref. \citen{klosigpic}, shown in fig. \ref{fig:sighad}, prove that taking advantage of ISR to effectively perform a scan in $s({\rm hadrons})$, at fixed collider energy is very effective and statistically powerful.
\subsection{KLOE results, what have we learned}
\begin{enumerate}
\item A \ff\ is a good source of physics, even with low \L\ and very large background
\item \DAF, with KLOE, has been a valuable venture, producing many highly skilled young people
\item LNF and INFN have profited from it and, for another couple of years will continue to do so, in a world which is becoming less and less sympathetic to research in fields remote from everyday connections. Which should never be a consideration\dots
\end{enumerate}
But the end is close, very close. We should find a way to go further.
\section{ A New \ff\ Program}\noindent
A \ff\ provides pure, monochromatic, low $\beta$ \ks, as well as \kl\, $K^+$ and $K^+$ beams with automatic absolute count. If we can do exceptional physics with the above boundary conditions it is justified to have a new \DAF. Before discussing what can be done, it is probably better to clear up some misconception about things that cannot be done there.

The process $\kl\to\po\nu\bar\nu$ cannot be studied. Since very wrong statements, see ref. \citen{fb} have been made and repeated about what the signal at a \ff\ might be, I must clarify this. Let us deal first with integrated luminosity and begin with \L=10\up{12} $\mu$b\up{-1} (=10\up6 1/pb or 1000 1/fb). With that much integrated \L, 10\up{12}\ks\kl\ pairs are produced. Tag is essential and one pays a factor 0.68 for BR(\ks\to\pic) and another factor 0.57~\cite{klopipibr} for reconstructing the \ks\to\pic\ decay. Finally, the fraction of \kl-meson decaying between 50 and 150 cm is 0.21. Put al this together with BR($\kl\to\po\nu\bar\nu$)=\pt2.6,-11,: 10\up{12}\x\pt2.6,-11,\x0.68\x0.57\x0.21=2.1 events. The final signal will be 2.1\x\eps, where \eps\ is the efficiency after applying a series of cuts to separate a very unclear signal from overwhelming backgrounds. The BNL experiment achieves \eps=\pt2,-3, for $K^\pm\to\pi^\pm\nu\bar\nu$. If you are a dreamer you might hope for \eps=1/10. So 0.2 events is the net result. Assuming that a super \ff\ delivers its average $\overline{\L}$ for 10\up{7} seconds, you find that for $\overline{\L}$=10\up{35} cm\up{-2} s\up{-1}, or \L\dn{\rm peak}=\pt2,35,, you have 0.21 events and lots of background.

The point is that under the most optimistic assumption, one needs one year of continuous running at the nominal, never achieved, efficiency of 33\%, at a luminosity of 10\up6 $\mu$b\up{-1}/s (10\up{36} cm\up{-2} s\up{-1}, not 10\up{35}) to collect 2 example of the $\kl\to\po\nu\bar\nu$ decay and an unknown number of background events. A real measurement of $\eta$ requires at least ten years or ten times higher luminosity. But our machine colleagues think that one tenth is already a dream. The case for $K^\pm\to\pi^\pm\nu\bar\nu$ appears better at first look, but the background problem is certainly worse, see ref. \citen{lp}.

The attitude that maybe one can find a trick to optimize the $\pi\nu\bar\nu$ search can be quite counterproductive. The important point is to have a program, not to do an experiment. This is clearly seen in the decisions in other labs and would not be good for LNF. In this respect, the ``60\deg'' collider, ref. \citen{panta}, is clearly counterproductive. It voids the unique advantages of a \ff, the \ks\ beam. A $\gamma\beta$ factor of \x6-8 is irrelevant for the \ks\ mean decay distance, but precision measurements of the \ks\ decay products become impossible. For the \kl\ decay, a detector \ab8 times longer than KLOE becomes necessary. That does not fit in the lab. Finally it must guarantee a factor four, or more, increase in luminosity to break even.

There is no question that if such measurements will ever be done, it will be at a hadron machines, where kaon yields 500-1000 times larger can be contemplated, not to say used. 
And maybe it is not so important. Before the outstanding success of the $B$-factories I used to think that a direct measurement of $\eta$ was a must. Now we know that $B$-meson can contribute, we even know that once more the SM model seems to work fine. Surprises might be just around the corner but that does not constitute a program. Furthermore, even if supersymmetry were to be proven to exist, it might very well be that nothing changes in the kaon sector, maybe not even in the whole CKM sector. At least at levels to which we might be sensitive in the next decade. So we leave that to \DAF3. If it were to happen that the rate for \ks\to\po$\nu\bar\nu$ is 100 times the expected value, we better always keep our eyes open. It is however very unlikely in the light of all we have learned in the past few years.

With that off my chest, lets examine what physics can we do. While \L=1000 nb\up{-1}/s is out of the question, \L=50 nb\up{-1}/s is conceivable, maybe some time down the road. It is a pity that a series of circumstances did not allow \DAF\ to resume running, after all the work done during January-June '03. With 50  nb\up{-1}/s, the \ks\ yield is \pt5,11, per year. Many things become interesting. In the SM \ks\to\pio\po, \to\po e\up\pm$\nu$ or the semileptonic asymmetry are trivially calculable from \kl\ BR's and \eps:\\[2mm]
1. BR(\ks\to\pio\po)=\pt1.9,-9, {($\pm$2.4\%)}\\
2. BR(\ks\to$\pi^\pm e^\mp\nu$)=\pt6.7,-4, {($\pm$1.5\%)}\\
3. ${\cal A}^\ell_S$=2$\Re\eps$=\pt3.323,-3, {($\pm$1.7\%)}\\
\dots\\
When something is so precisely predicted, it sort of becomes a must to measure it.
We shall ignore the usual comment of how much discovery range is avilable\dots

\begin{figure}[htb]
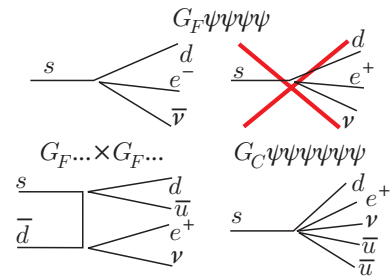

\centering
\figboxc dsdqnot3;5;
\caption{Amplitudes for $\Delta S=\Delta Q$ and apparent $\Delta S=-\Delta Q$ transitions.}
\label{fig:dsdqnot}
\end{figure}

The above ideal \ff\ with the given BR's means the production of
$$\vbox{\halign{\hfil#&\quad#\cr
$N(\ks\to\pio\po)$&950/y\cr
$N(\ks\to\pi^\pm e^\mp\nu)$&\pt3.3,8,/y\cr}}$$
Not all decays can be collected, but at least all \ks's decay in the detector. Also 950 is a large enough number, even with discounts on the luminosity and after cuts to suppress background. And finally one can collect data for a few years.
For comparison, NA48 collected \pt5,6, \kl\to\pio\ decays between 1997 and 2001, the original proposal having been submitted in 1990. See Ceccucci, this workshop. Anyway we see the beginnings of a program.
There are many ingredients and parameters of the standard model about which we do not know nor do much. For instance $\Delta S=\Delta Q$. At the quark level there are no $\Delta S=-\Delta Q$ transitions, which, see fig. \ref{fig:dsdqnot}, are tantamount to violating charge conservation.
Introducing the standard ratio of amplitudes $x=A(\ko\to e^+)/A(\kob\to e^+)$ we can estimate, from the second order diagram of fig. \ref{fig:dsdqnot}, $x\ab Gm^2\ab10^{-6\ {\rm or}\ -7}$. Or from compositeness, last graph, $x=10^{-10}\x({\rm1\ GeV}/\Lambda)^6$. But notice there are no loops, which means you can't mess up the estimate with supersymmetry. There are in general two $x$'s: $x^+, x^-$. Experimentally we know $\Re x<0.1\%$. We must do better. Another job for \DAF2.

The quark mixing matrix, $V_{\rm CKM}$ must be unitary. Every now and then there is an uproar, because the sum of the moduli of the first row elements misses 1 by maybe 2\sig. Remember however how strong are the constraints on unitarity from $\Delta M_K$ and BR($\kl\to\mu\mu$). What is to me truly surprising is that $|V_{us}|$ is known to no better than 1.5\%. 3\sig\ effects have been known to disappear, 2\sig\ is no worry but 1.5\% error on a parameter that we have been trying to measure for 40 years, since 1963, is scandalous. Maybe the extraction of $|V_{us}|$ from quantities such as $\Gamma(K\to\pi\ell\nu)$ will always be difficult but we ought to measure the widths better. By the way if they where truly well measured we could check the calculations of many people by comparing widths for neutral and charged kaons. I really believe that continuing to reduce the errors on these widths is a truly noble endeavor.

So maybe we can postpone checking CKM, meanwhile measuring better $\lambda$, which is necessary also to unravel the information from $\K\to\pi\nu\bar\nu$. That requires knowledge of the fifth power of $\lambda$, which today has an error of 5\x1.5=7.5\%. We can leave the $\pi\nu\nu$ decays to \DAF3\ and measure semileptonic widths at \DAF2.
Remember that it took 40 years to get from the discovery of \nocp\ to the present relatively good value for \rep. And it was done by NA48-KTeV -- but -- after the experience of NA31-E731. So it took 20 years! The same was true about the two $K^+\to\pi^+\nu\bar\nu$ events at BNL.

Direct \nocp\ can also be searched for in \K\up\pm\ decays, measuring rate and slope asymmetries. The NA48/2 effort could be continued at a new \ff, possibly reaching better sensitivity. Remember $A_g\ab10^{-6}$ and  $A_\Gamma\ab10^{-8}$ and not even the authors, GGG as explained in ref. \citen{ggg}, like the supersymmetric enhancement via a ``chromo-magnetic'' operator ({\it possible only if\dots several conditions\dots conspire}, their words) which they invoke.

\subsection{\C\P\T}\noindent
In 1957-64 we saw the demise of \P, \C\ and \C\P\, what about \C\P\T? This is a most important reason for studying \ks\ decays. Let me notice right away that we must aim for ${\cal O}$(10\up{-5}) sensitivity or 10\up{9} semileptonic \ks\ decays.

The kaon system does provide the strongest upper bound on $\Delta M/\langle M\rangle$ for \C\P\T\ conjugate states. Of course since we do not really know what to expect, we do not know when we have achieved a significant --null-- result.
An argument made in the past is that one should compare the dimensionless ratio $\Delta M_K/M_K$ with another dimensionless ratio $M_K/M_{\rm Plank}$=0.5/\pt1.2,19,\ab\pt4,-20,. There one might contemplate loss of QM coherence or non flat space, thus losing the bases for the Pauli-L\"uders theorem. We should therefore aim for  $\Delta M_K$\ab\pt2,-11, eV.
Without assuming \C\P\T\ invariance, to l.o. in ``\eps''~\cite{pfa}:
$$\eqalign{
\sta{\ks}&=[(1+\epss)\sta{\ko}+(1-\epss)\sta{\kob}]/\sqrt 2\cr
\sta{\kl}&=[(1+\epsl)\sta{\ko}+(1-\epsl)\sta{\kob}]/\sqrt 2\cr
}$$
The variables $\tilde\eps$ and $\delta$ defined through the identities
$$\epss\equiv\tilde\eps+\delta\qquad\epsl\equiv\tilde\eps-\delta.$$ 
are illustrated in fig. \ref{fig:etarho}. Using unitarity, ${\cal A}^e_L$, etc. and assuming no \nocpt\ in the decay amplitudes leads to limits on $\delta$ and
$${|M(\ko)-M(\kob)|\over \langle M\rangle}={\Delta M\over M}=(2\pm9)\x10^{-19}.$$
Without any assumption about \nocpt, or $\Gamma(\ko)=\Gamma(\kob)$, the result is considerably weaker, \ab few\x10\up{-18}. From ref. \citen{pfa},
$$|M(\ko)-M(\kob)|=|\Gamma_S-\Gamma_L|\,|\Re\delta\tan\phi_{SW}-\Im\delta|.$$
We therefore need measuring $\delta$ to 2/3\x10\up{-5}.
In general, but with $\Delta S=\Delta Q$, $\Re\delta=({\cal A}^e_L-{\cal A}^e_S)/4$. Therefore we need to reach an accuracy in ${\cal A}$ of \ab\pt3,-5, or collect 10\up9 events. 

\section{KLOE at \DAF2}\noindent
There is a long list of interesting things to to, if \DAF2\ can reach \L=50 nb\up{-1}/s.
There are still unobserved decay modes such as \ks\to\pio\po. And to push the measurement of $|\Delta S|$ to the level necessary for reaching 0.01\% accuracy in $|V_{us}|$ is also worthwhile. In addition there is the following.
\begin{Enumerate}
\item Continue the original KLOE program on direct \nocp, precision measurements of decay rates, hadronic cross section, scalar and pseudo scalar mesons, semi-rare decays.
\item A direct study of $\Delta S=\Delta Q$. For this we can use charge exchange, $K^+\Rightarrow\ko,\ K^-\Rightarrow\kob$ to tag strangeness by strong interactions, thus avoiding ambiguities from possible \C\P\T\ violation.
\item Use interference to measure to measure magnitude and phase of the amplitude ratios $\eta_i$: $\Re\eta_{+-},\ \Im\eta_{00}$ up to, $\Im\delta$
\item Study \ks\ and \kl, especially the leptonic asymmetry to reach enough accuracy on $\Re\delta$ or M(\ko)-M(\kob).
\item Slope and rate asymmetries in $K^\pm$ 3 pion decays, to better than 10\up{-4}.
\item Push all accessible modes to the limit of \ab 10\up{-11}.
\end{Enumerate}
It is a long program, especially since overall efficiencies are not 1, but they are not 0.01 either. There should improvements in KLOE, after all \kl\ was at premium in its original design, but \ks\ dominates now. But that belongs to another discussion

There is one crucial point that must be reaffirmed. Higher luminosity must come together with low background, less than now. Moreover is the total background that matters not the ratio of background to luminosity. An increase of luminosity of a factor 100 together with a background increase of 100 would make it impossible to do any of the physics discussed. Even more a \x100 increase in luminosity, without a \x10 decrease in background would also be useless. This is often not appreciated.

\section{CONCLUSIONS}\noindent
KLOE still hopes to collect $\int\L\dif t>$1 fb\up{-1} in 2004, to complete the first phase of a successful program.

Beyond that, the \DAF2 collider discussed could ultimately lead to a \x100 increase in \L. Lower \L\ still ensures a very exciting physics program, to be well underway before the end of this decade. A new, improved \ff-collider could lead to exciting results for a period of 5 to 10 years.

The SM fares extremely well at LEP, SLAC, the Tevatron and even at BNL with the muon $(g-2)$ measurements, but we do not know have critical tests of the validity of the $\Delta S=\Delta Q$ rule, about CKM unitarity, invariance under $CPT$\dots

\bye